# Anderson localization in Bragg-guiding arrays with negative defects

Valery E. Lobanov[1], Yaroslav V. Kartashov[1,2], Victor A. Vysloukh[3], and Lluis Torner[1]

[1]ICFO-Institut de Ciencies Fotoniques, and Universitat Politecnica de Catalunya, 08860 Castelldefels (Barcelona), Spain
[2]Institute of Spectroscopy, Russian Academy of Sciences, Troitsk, Moscow Region, 142190, Russia
[3]Departamento de Fisica y Matematicas, Universidad de las Americas—Puebla, 72820 Puebla, Mexico
*Corresponding author: valery.lobanov@icfo.eu

We show that Anderson localization is possible in waveguide arrays with periodically-spaced defect waveguides having lower refractive index. Such localization is mediated by Bragg reflection, and it takes place even if diagonal or off-diagonal disorder affects only defect waveguides. For off-diagonal disorder the localization degree of the intensity distributions monotonically grows with increasing disorder. In contrast, under appropriate conditions, increasing diagonal disorder may result in weaker localization.

OCIS Codes: 190.5940, 130.2790, 240.6690

Anderson localization was initially discovered in solid state physics [1,2] upon analysis of localization of electronic wave functions in disordered crystals, but it was soon realized that it occurs in several areas of science, including acoustics [3], physics of microwaves [4], matter waves [5], and optics [6-10]. Disordered optical waveguide arrays provide a unique laboratory for the investigation of such phenomena, because a suitable longitudinally invariant disorder required for transverse Anderson localization (rapid longitudinal disorder variations may destroy the effect) is readily realizable in such structures [10-12]. Thus, transverse Anderson localization has been observed in optically induced [11] and fabricated lattices [12-16], in both two- [11,15,16] and one-dimensional [12-14] settings. The effects of disorder inhomogeneity [17], nonlinearity [12,18,19] and array interfaces [13,16,19,20] have been also analyzed.

However, in all previous studies disorder-induced localization has been studied in periodic systems, which for vanishing disorder consist of identical waveguides, where light localization is achieved due to total internal reflection. The possibility of Anderson localization in systems with Bragg-guiding mechanism has not been addressed so far.

In this Letter we show that Anderson localization can be observed in defect arrays, in which light trapping on periodic defects with reduced refractive index is stimulated by Bragg-type reflections [21-26]. We find that in such structures, localization is possible even when diagonal or off-diagonal disorder is introduced only into defect waveguides.

We address light propagation in one-dimensional disordered Bragg-guiding arrays described by the nonlinear Schrödinger equation for the normalized field amplitude $q$:

$$i\frac{\partial q}{\partial \xi}=-\frac{1}{2}\frac{\partial^2 q}{\partial \eta^2}-R(\eta)q+\sigma q|q|^2. \quad (1)$$

Here $\eta$ and $\xi$ are the normalized transverse and longitudinal coordinates, respectively; $\sigma=\mp 1$ corresponds to focusing/defocusing nonlinearity (note that except in Figs. 5(a) and 5(b), we consider linear propagation thus $\sigma = 0$); the periodic array $R(\eta)=\sum_{k=-\infty}^{+\infty} p_k G(\eta-\eta_k)$ is formed by super-Gaussian waveguides $G(\eta)=\exp(-\eta^6/w_\eta^6)$ with width $w_\eta$ and refractive index $p_k$. Defect waveguides with refractive index $p_k=p_\mathrm{d}$ are periodically nested into the array and are separated by $m$ waveguides with higher refractive index $p_k=p_\mathrm{a}$. In the absence of defects, i.e. at $p_\mathrm{d}=p_\mathrm{a}$, the linear modes $q(\eta,\xi)=w(\eta)\exp(ib\xi+i\kappa\eta)$ of Eq. (1) are Bloch waves, with $w(\eta)=w(\eta+d)$, $d$ being the waveguide spacing and $\kappa$ the Bloch momentum.

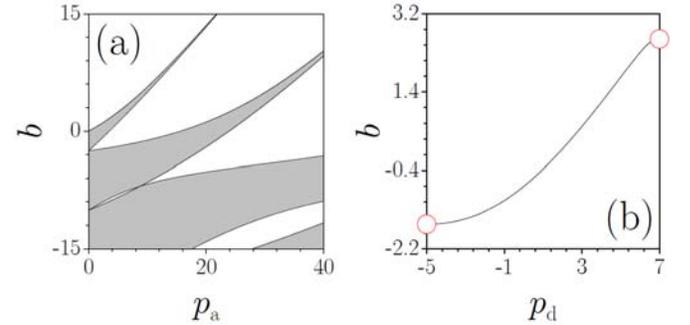

Fig. 1. (a) Band-gap spectrum of a uniform lattice. (b) Propagation constant of the defect mode versus $p_\mathrm{d}$ for $p_\mathrm{a}=7$. Red circles indicate gap borders.

The propagation constants $b$ form bands [gray regions in Fig. 1(a)] separated by gaps (white regions). Localized linear defect modes appear when $p_\mathrm{d} \neq p_\mathrm{a}$ at least in one channel. If $p_\mathrm{d} < p_\mathrm{a}$ such modes form due to Bragg-type reflections - they have oscillating tails and propagation constants $b$ belonging to the gaps. The modes with $b$ values in the middle of the gap are well localized, their counterparts with $b$ values close to the gap edges are extended. In the case of a single defect, the propagation constant of the defect mode approaches the upper gap edge at $p_\mathrm{d} \to p_\mathrm{a}$, while with decreasing $p_\mathrm{d}$ it approaches the lower gap edge [Fig. 1(b)]. If there is a periodic sequence of defects in the array, one can observe tunneling of light between the defects, which is linked with coupling of defect modes that results in the expansion (transport) of the input excitations.

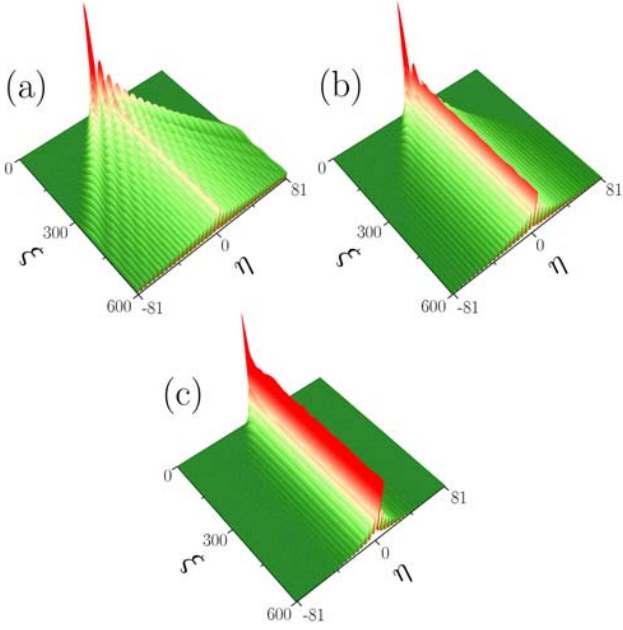

Fig. 2. Dynamics of light propagation in disordered lattices with negative defects, averaged over $Q=10^3$ disorder realizations. A disorder with $S_d = 0.15$ (a), $0.25$ (b), and $0.4$ (c) is introduced into positions of the defect channels. In all cases: defect waveguides are separated by two waveguides; $p_d = 4$, $p_a = 7$ and $\sigma = 0$. The plots show the field modulus.

To study Anderson localization we use arrays in which disorder is introduced into defect waveguides only. In the case of off-diagonal disorder [13] the coordinates of the defect waveguide centers $\eta_k = (m+1)dk + r_k$ are randomized, with $d$ being a regular spacing and $r_k$ being a random shift of the waveguide position uniformly distributed within the segment $[-S_d, +S_d]$, while the positions of the ordered waveguides are fixed. If diagonal disorder is considered, only the depths of the defect waveguides are randomized, as $p_k = p_d + n_k$, with $n_k$ being a random refractive index perturbation uniformly distributed within $[-S_d, +S_d]$. For each set of parameters we generated $Q \sim 10^3$ realizations of disordered arrays and solved Eq. (1) up to large propagation distances ($\xi \sim 600$) for each disorder realization. As input conditions we used $q|_{\xi=0} = A w_d(\eta)$, where $w_d$ is the linear mode of an array with a single defect with amplitude $A$. For statistical analysis the intensity distributions were averaged over the ensemble of array realizations $I_{av} = Q^{-1}\sum_{i=1}^Q |q_i|^2$, while light localization was characterized by the averaged integral form-factor $\chi_{av} = Q^{-1}U^{-2}\sum_{i=1}^Q \int_{-\infty}^{+\infty} |q_i|^4 \, d\eta$, where $U = \int_{-\infty}^{+\infty} |q_i|^2 \, d\eta$ is the energy flow, which was constant for each realization. Notice that the inverse integral form-factor $\chi_{av}^{-1}(\xi)$ characterizes the width of the core of the averaged intensity distribution, almost ignoring a low-intensity background.

We start from the case of off-diagonal disorder in the linear limit ($\sigma = 0$) and set $d = 1.4$ (which under standard conditions correspond to a waveguide spacing equal to 14 μm), $w_\eta = 0.3$ (waveguide width equal to 3 μm), $p_a = 7$ (refractive index modulation $\sim 7.8 \times 10^{-4}$ at $\lambda \sim 800$ nm [24]). We assume that defect waveguides are separated by $m = 2$ waveguides. Figure 2 shows the ensemble-averaged field modulus distributions $|q(\eta, \xi)|_{av} = Q^{-1}\sum_{i=1}^Q |q_i|$ for different disorder levels $S_d$. The plots show how by increasing the disorder level $S_d$ transport in regular arrays with periodic defects is replaced by the formation of strongly localized averaged distributions, which remain virtually unchanged after propagation. Thus, Anderson localization occurs even in systems with Bragg-guiding mechanisms, where light concentrates in low-index regions. Anderson localization is also readily apparent in Figs. 3(a) and 3(b) that illustrate the details of the averaged output intensity distributions in a logarithmic scale. At high disorder levels nearly perfect triangular distributions form, indicating an exponential localization of their tails. Notice the presence of additional small maxima around the centers of the waveguides that are located between the main spikes, whose positions coincide with the average positions of defect waveguides in Figs. 3(a),(b). The localization distance [i.e. the distance at which $\chi_{av}^{-1}(\xi)$ reaches its steady-state value] rapidly decreases by increasing the disorder level $S_d$ [Fig. 3(c)]. The width of the averaged output distribution, characterized by $\chi_{av}^{-1}$, monotonically decreases when increasing $S_d$ [Fig. 4(a)]. We also compared the cases when off-diagonal disorder affects only the positions of the defect waveguides and when disorder is introduced into positions of all waveguides. As expected, in the second case the output distributions exhibit a higher localization, especially for low disorder levels [Fig. 4(a)].

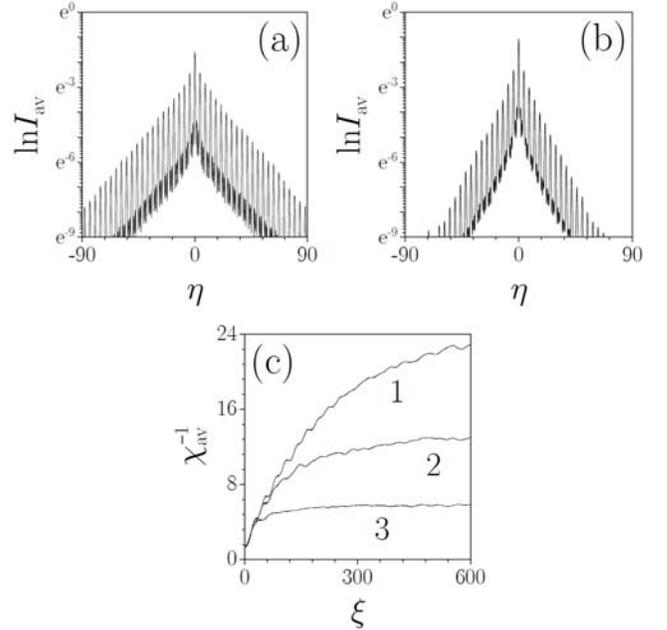

Fig. 3. Averaged output intensity distributions at $\xi = 600$ for (a) $S_d = 0.25$ and (b) $S_d = 0.35$. (c) Averaged inverse form-factor versus propagation distance at $S_d = 0.15$ (curve 1), $S_d = 0.2$ (curve 2), and $S_d = 0.3$ (curve 3). In all cases $p_d = 4$, $p_a = 7$, $\sigma = 0$.

One of the key parameters determining the rate of tunneling between neighboring defects in Bragg-guiding array is the defect depth $p_d$. Surprisingly, at a fixed disorder level, the output averaged form-factor $\chi_{av}$ turns out to be a non-monotonic function of $p_d$ [Fig. 4(b)]. Thus the strongest localization occurs when isolated defects support strongly localized modes with $b$ values inside the gap. In such a regime, the tunneling rate between neighboring defects is minimal and even a weak disorder is sufficient to generate a strong localization. In contrast, at $p_d \to p_a$ the defect modes are weakly localized, the coupling between defects increases, and localization diminishes for a fixed value of $S_d$. At $p_d \to 0$, localization decreases even faster because a reduction of $p_d$ leads to a decreasing disorder strength. A behavior similar to that shown in Fig. 4(b) was obtained for Gaussian beams.

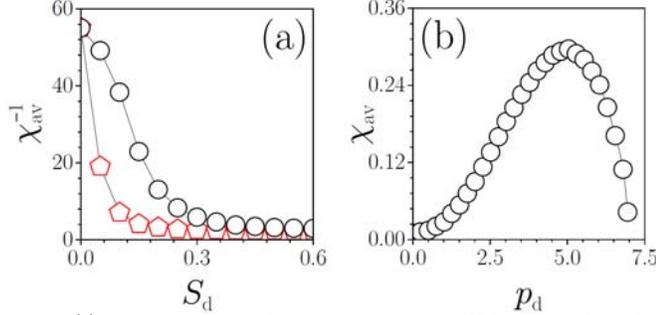

Fig. 4. (a) Inverse averaged form-factor at $\xi=600$ versus disorder strength for $p_d=4$, when disorder affects only the positions of the defect waveguides (circles) and when disorder affects the positions of all waveguides (pentagons). (b) Averaged form-factor at $\xi=600$ versus defects depth for $S_d=0.4$. In all cases $p_a=7$, $\sigma=0$.

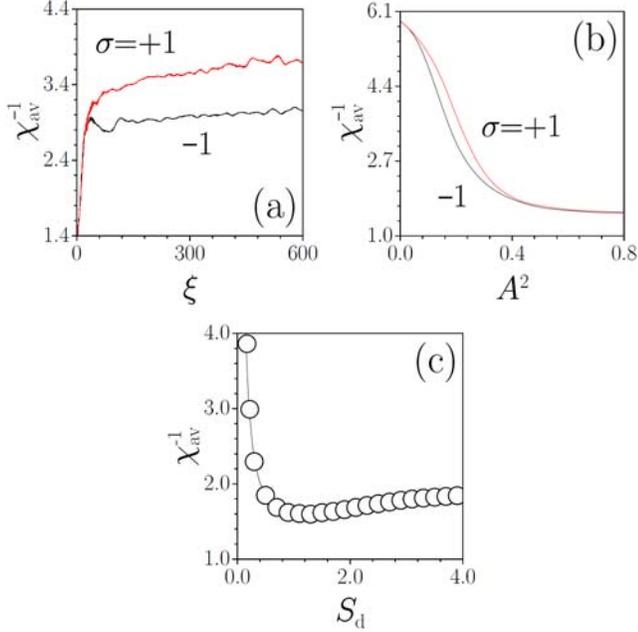

Fig. 5. (a) Inverse averaged form-factor versus propagation distance at $A^2=0.2$ in focusing ($\sigma=-1$) and defocusing ($\sigma=+1$) media. (b) Inverse averaged form-factor at $\xi=600$ versus input peak intensity. In panels (a),(b) $p_d=4$, $p_a=7$, $S_d=0.3$. (c) Inverse averaged form-factor versus strength of diagonal disorder at $\xi=1000$ for $p_d=5$, $p_a=7$, $\sigma=0$.

We observed that in the Bragg-guiding arrays considered here, both focusing and defocusing nonlinearities result in enhancement of localization [see Figs. 5(a) and 5(b)], even for small input amplitudes $A$. In the focusing case localization is more pronounced. As expected, at large amplitudes disorder-induced localization is gradually replaced by bright soliton formation in focusing media and by gap soliton formation in defocusing media.

Finally, we found that diagonal disorder also results in Anderson localization. In contrast to the case of off-diagonal disorder, diagonal disorder generates a non-monotonic dependence of the averaged inverse form-factor $\chi_{av}^{-1}$ versus the disorder level $S_d$ [Fig. 5(c)]. Such dependence is related to the behavior of the averaged field distributions at large disorder levels $S_d \geq |p_a - p_d|$. While for small $S_d \sim 0.5$ the main spikes in $I_{av}(\eta)$ are located near the defect waveguides, like in Fig. 3(a), at $S_d \sim 2.0$ light starts concentrating also on ordered waveguides, resulting in $I_{av}(\eta)$ distributions without small-amplitude oscillations.

Summarizing, we found that Anderson localization is possible not only in systems where guidance is mediated by total internal reflection, but also in Bragg-guiding arrays. In such arrays the localization degree varies non-monotonically with the depth of the defect waveguides.

# References with titles